\documentclass[lettersize,journal]{IEEEtran}
\usepackage{amsmath,amsfonts}
\usepackage{algorithmic}
\usepackage{algorithm}
\usepackage{array}
\usepackage{textcomp}
\usepackage{stfloats}
\usepackage{url}
\usepackage{cite}
\usepackage{verbatim}
\usepackage[pdftex]{graphicx}
\usepackage{color}
\usepackage{transparent}
\usepackage{import}
\usepackage{subfigure}
\usepackage{balance}
\usepackage{multirow}
\usepackage{cleveref}
\DeclareGraphicsExtensions{.eps}
\usepackage{epstopdf}
\usepackage{epsfig}
\usepackage{gensymb}
\usepackage{mathrsfs,amsmath}
\usepackage{float}
\usepackage{amsmath,amsfonts,amssymb}
\begin{document}


\title{Millimeter-Wave Dual-Polarized Omnidirectional Reference Antennas for Total Array Gain Evaluation}

\author{Bing Xue,~\IEEEmembership{Member,~IEEE}, Juha Tuomela, Katsuyuki Haneda,~\IEEEmembership{Member,~IEEE}, and Clemens Icheln
\thanks{Manuscript received July XX, 2025; revised August XX, 2025.}
\thanks{The authors are with the Department of Electronics and Nanoengineering, Aalto University-School of Electrical Engineering, Espoo FI-00076, Finland (e-mail: bing.xue@aalto.fi).

Y. Yuan is also with the State Key Laboratory of Advanced Rail Autonomous Operation, the School of Electronics and Information Engineering, and the Frontiers Science Center for Smart High-speed Railway System, Beijing Jiaotong University, Beijing 100044, China
}}
\markboth{Journal of \LaTeX\ Class Files,~Vol.~14, No.~8, July~2025}%
{Shell \MakeLowercase{\textit{et al.}}: A Sample Article Using IEEEtran.cls for IEEE Journals}

\maketitle
\begin{abstract}
The present manuscript introduces a compact millimeter-wave dual-polarized reference antenna module designed for performance comparison of handset antenna arrays in multipath environments. Both vertically- and horizontally-polarized fields are covered with sufficiently wide impedance bandwidths and low gain variation along the horizontal plane. When integrated into a compact dual-polarized antenna module, the mutual coupling between the antennas is well controlled to ensure minimal distortion in the radiation patterns. Measurements confirm that the proposed antenna module achieves the designed low gain variation and almost identical realized gains for the two polarizations.
\end{abstract}
\section{Introduction}
\IEEEPARstart{T}{he} fifth-generation (5G) radios have been deployed widely, but mainly at sub-6 GHz bands~\cite{Bae2025A}. While these frequencies offer good coverage and penetration, their limited bandwidth constrains achievable data rates~\cite{Chen2019A}. To meet growing demands for ultra-high-speed manuscript, future 5G advancements and sixth-generation (6G) systems are expected to utilize millimeter-wave (mmWave) and higher frequency bands more aggressively~\cite{Guan2021Channel,Patrick2025Automated,Bae2025A,Wang2024Sub}. 

Handset antennas, as integral components of the wireless link, have been extensively investigated in mmWave systems. Accurate performance evaluation is essential for reliable antenna design, often requiring reference antennas with well-characterized radiation patterns. Depending on the application, different reference antennas are used. Standard gain horns are commonly employed in antenna characterization~\cite{Chen2006The,Patel2018A,razmhosseini2020accuracy}; directive patch arrays are preferred in user blockage studies due to their beamforming capabilities and low cross-polarization~\cite{Lauri2021,Xue2023Impacts,Xue2025Handset}.
Omnidirectional antennas are also needed when comparing handset arrays in multipath environments~\cite{Haneda2018,Haneda2018_2}. As with anechoic measurements, dynamic performance evaluations require reference antennas with stable and known patterns~\cite{Suzan2019Design}. However, unlike the chamber environment, reference antennas for realistic channel conditions must be omnidirectional, dual-polarized, and exhibit minimal gain variation in the horizontal plane. Moreover, to investigate dynamic fading channels, both polarizations must be measured simultaneously at the same location, posing notable design challenges~\cite{sulonen2004evaluation}. 

While many omnidirectional antennas have been developed for sub-6 GHz applications~\cite{balanis2016antenna,Suzan2019Design,Zhou2021A,Liang2022Compact,Zhang2017Bandwidth,Lin2019A,Zhang2019Wideband,Dobler2020An,Ratajczak2019Design}, mmWave designs remain scarce due to several difficulties: 1) coaxial connectors become comparable in size to the antenna, disturbing the radiation pattern; 2) impedance characteristics at mmWave frequencies are highly sensitive to fabrication tolerances~\cite{Xue2025Handset}; and 3) reference antennas must ensure minimal gain variation across the horizontal plane~\cite{Haneda2018,Haneda2018_2}. Although~\cite{Zhai2024A,Yan2024A,Zhou2025A,Ke2025Dual} have proposed dual-polarized omnidirectional antennas, their monopole-based structures are unsuitable for mmWave applications due to severe cable-induced pattern distortion, particularly for vertically polarized (V-pol) elements. Moreover, the feed cables are typically too bulky to support both V-pol and H-pol excitations within a shared aperture. All of the above challenges restrict the development of dual-polarized omnidirectional antennas at mmWaves.

To address these challenges, this manuscript presents a practical omnidirectional reference antenna module for mmWave channel measurements. Two orthogonally polarized antennas, i.e., V-pol and horizontally polarized (H-pol) antennas, are individually designed at 28~GHz. Using thin semi-rigid coaxial cables, metal painting, and conductive adhesives, both elements are fabricated with wide impedance bandwidths, low horizontal gain variation, and low manufacturing complexity. Integrated as a compact dual-polarized module, the antennas maintain low mutual coupling and minimal pattern distortion. The resulting design offers a practical reference solution for handset testing in realistic multipath environments.

\section{28 GHz H-pol Omnidirectional Antenna}
\label{sec:H-pol}
\subsection{Antenna Design}
\label{sec:H-poldesign}
The H-pol omnidirectional antenna design is primarily based on the concept of an equivalent electric current loop, where the antenna size is approximately half a wavelength to achieve low side lobes and controlled gain variation~\cite{balanis2016antenna}. Standard loop antennas implemented on PCBs are not suitable for this purpose because connectors located on the horizontal plane disturb the antenna's radiation pattern~\cite{Suzan2019Design}. Designs based on equivalent magnetic current sources also show horizontal-plane gain variations of $1.5$~dB~\cite{Zhou2021A,Liang2022Compact} due to their structural asymmetry. A classical approach to realize an equivalent electric current loop is to employ $N$ dipole elements excited by a 1-to-$N$ power divider. 
Typically, $N = 4$ is used~\cite{Zhang2017Bandwidth,Lin2019A}.

\begin{figure}[!ht]
\centering
	 \subfigure[]{\label{fig:designHpol1}
	\includegraphics[width=0.53\linewidth]{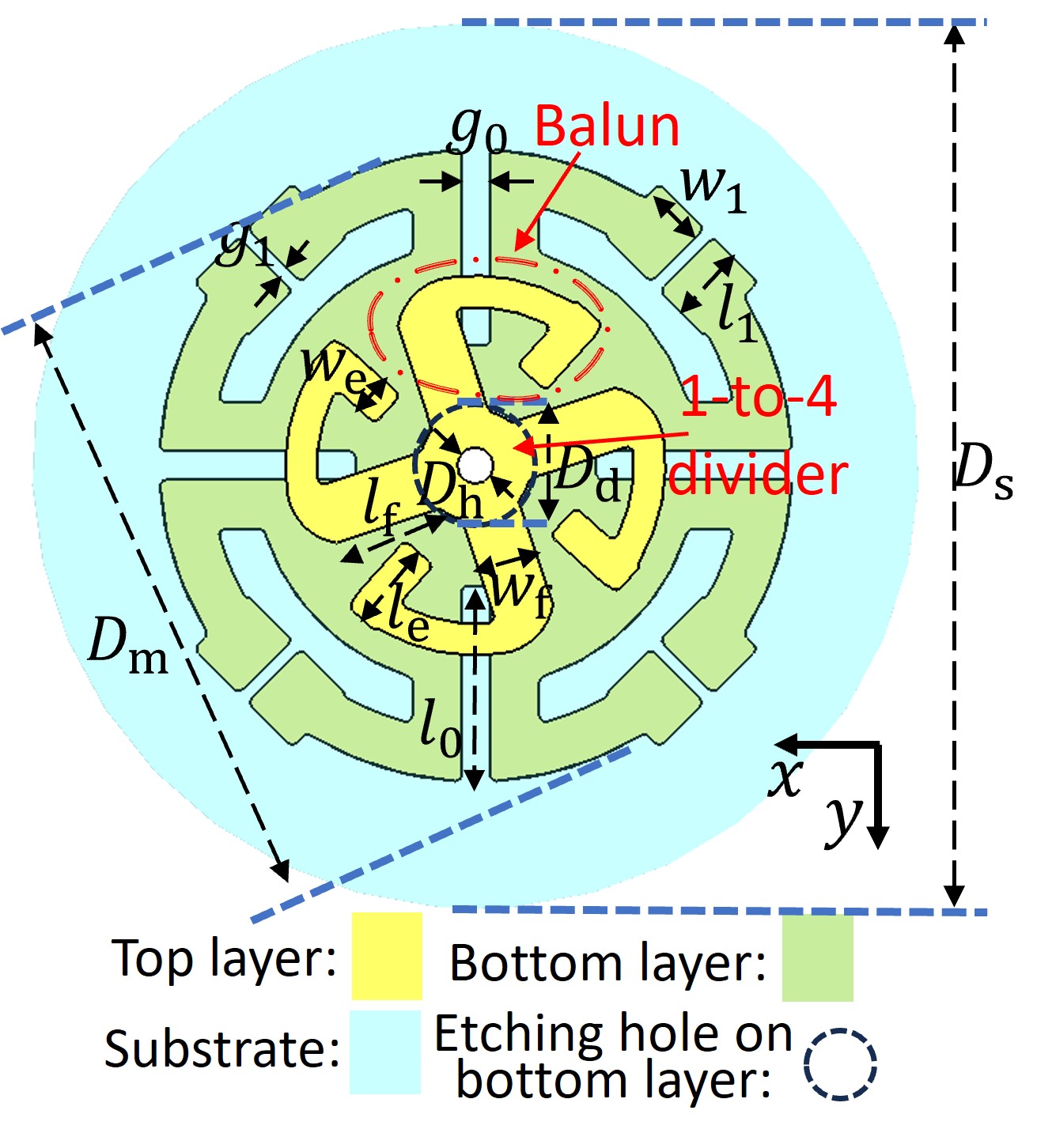}}
 	 \subfigure[]{\label{fig:designHpol2}
	\includegraphics[width=0.43\linewidth]{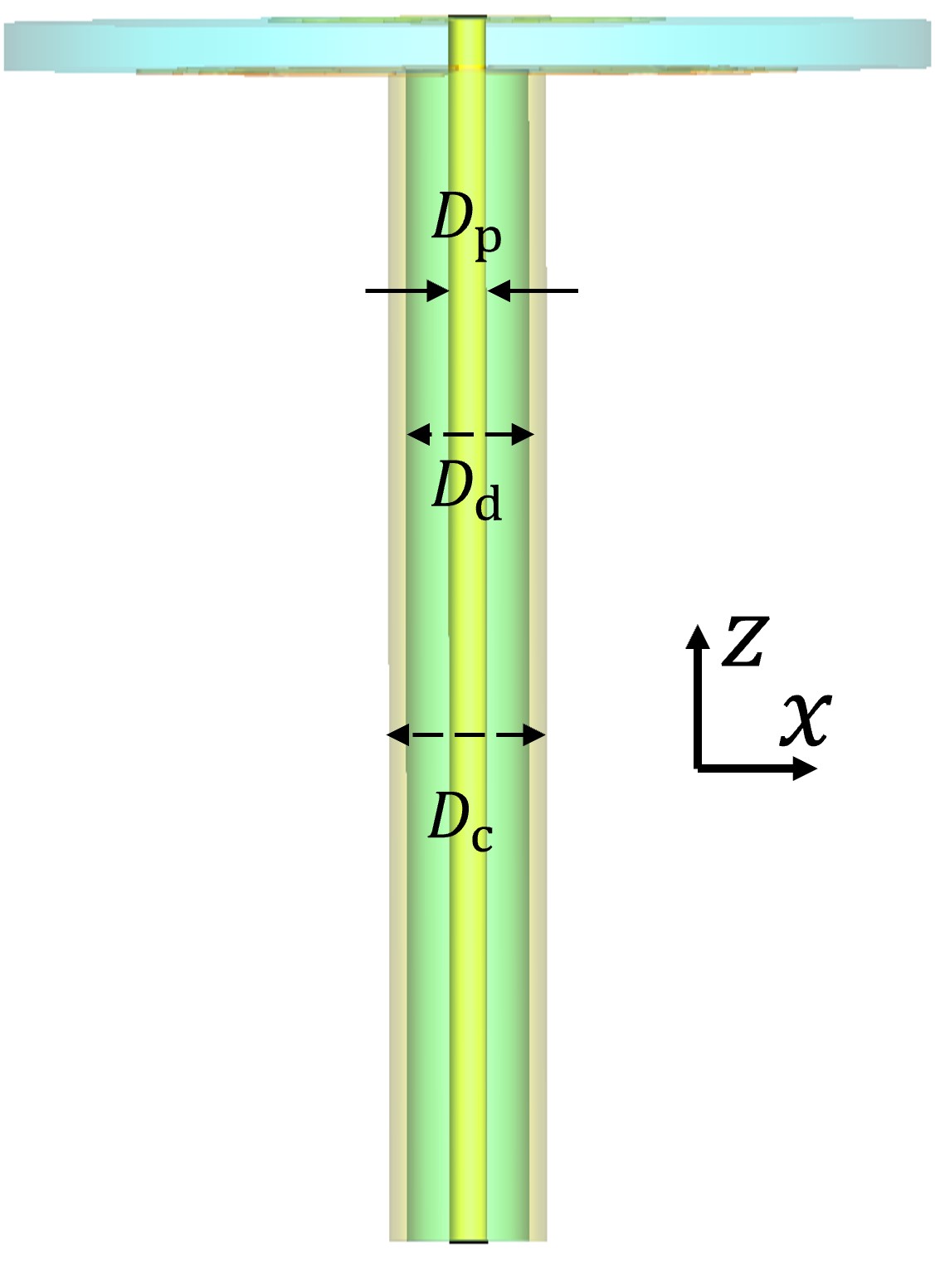}}
\caption{
(a) designed PCB of H-pol antenna; (b) designed H-pol antenna with feeding coaxial cable.}
	\label{fig:Hpoldesign}
\end{figure}

Unlike previously published H-pol omnidirectional antennas based on equivalent current loops, we adopt four compact, capacitively coupled dipole elements to reduce the overall loop size and reduce gain variation, as shown in Fig.~\ref{fig:designHpol1}. As the capacitive coupling among dipole elements increases, the required dipole length becomes shorter. To enhance coupling between adjacent elements, small patches ($l_1 \times w_1$) are introduced. Each of the four dipoles is excited via a slot transmission line. A balun is used to convert from unbalanced fields to balanced fields of Transverse Electromagnetic (TEM) mode, i.e., from the microstrip line to the slot transmission line. The inner conductor of the coaxial cable is connected to a 1-to-4 microstrip power divider, while the outer shield is grounded, as shown in Fig.~\ref{fig:designHpol2}. Given the working wavelength of approximately $10.7$~mm, the selected coaxial cable has a diameter of $D_{\rm c} = 1.19$~mm. The cable is a semi-rigid type with Teflon dielectric filling. In principle, due to the thin cable and the presence of baluns, the cable length does not significantly influence the antenna’s radiation pattern or resonance frequency. To minimize the microstrip width and enhance coupling within the balun, the substrate thickness should be as small as possible. However, soft or overly thin materials compromise mechanical strength and are avoided. Based on these considerations, we selected Megtron7 R-5785N (dielectric constant: 3.34, loss tangent: 0.003) with a thickness of $0.25$~mm. The metal layer thickness is $18~\rm \mu m$. The physical dimensions of the proposed antenna are listed in Table~\ref{tab:Hpolparameter}.
\begin{table}[!ht]
	\begin{center}
		\caption{Dimensions of the proposed H-pol omnidirectional antenna} 
		\label{tab:Hpolparameter}
        \begin{tabular}{llll}\hline\hline\multicolumn{4}{c}{H-pol omnidirectional antenna (unit: $\rm mm$)}\\\hline\hline
        $l_1 = 0.60$&$w_1 = 0.50$&$g_1 = 0.10$&$g_0 = 0.23$\\
        $l_0 = 1.50$&$l_{\rm e} = 0.76$&$w_{\rm e} = 0.30$&$l_{\rm f} = 0.83$\\
        $w_{\rm f} = 0.48$&$D_{\rm h} = 0.30$&$D_{\rm d} = 0.94$&$D_{\rm m} = 5.00$\\
        $D_{\rm s} = 7.00$&$D_{\rm p} = 0.29$&$D_{\rm c} = 1.19$&\\\hline
		\end{tabular}
	\end{center}
	\end{table}
 
\begin{figure}[!ht]
\centering
	 \subfigure[]{\label{fig:Hpolsparm}
    \includegraphics[width=0.48\linewidth]{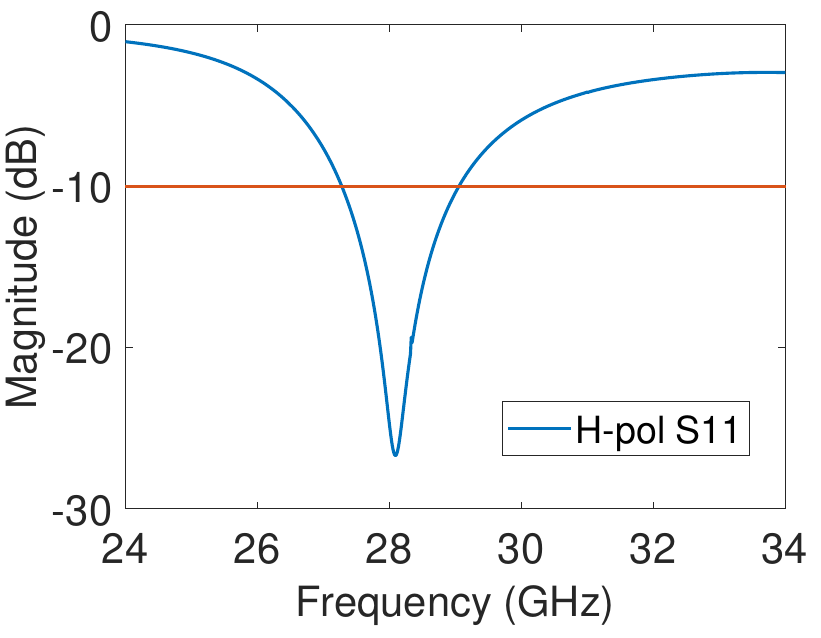}}
	 \subfigure[]{\label{fig:HpolEfield}
    \includegraphics[width=0.48\linewidth]{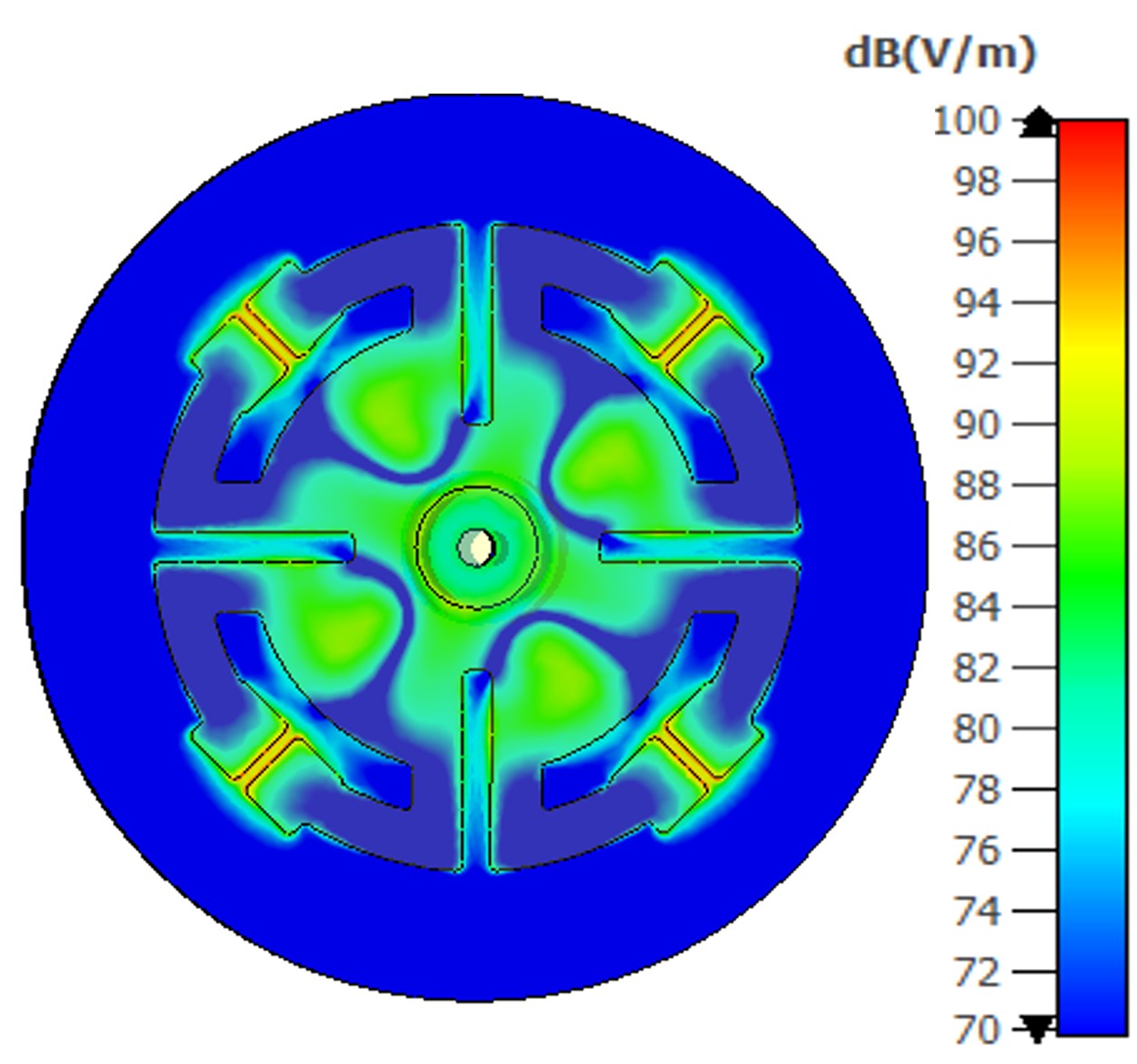}}
\caption{(a) Reflection coefficients, (b) E-Field on the substrate at $28$~GHz of the H-pol omnidirectional antenna.}
 \label{fig:Hpolsimulation}
\end{figure}

To minimize discrepancies between the simulation model and the fabricated prototype, practical etching effects at the PCB corners were incorporated into the simulation. Each corner was replaced by a circular chamfer with a radius of $50~\rm \mu m$ to reduce the manufacturing errors. The $10$~mm coaxial cable was excited using a waveguide port, and full-wave simulations were performed using \textit{CST~Studio~Suite}. The corresponding simulation results are presented in Fig.~\ref{fig:Hpolsimulation}. As shown in Fig.~\ref{fig:Hpolsparm}, the antenna resonates at $28$~GHz with an impedance bandwidth of approximately $1.7$~GHz. The electric fields (E-fields) on the substrate confirm that strong coupling occurs between adjacent dipoles, and the baluns effectively transfer power from the microstrip lines to the slot transmission lines.

\begin{figure}[!ht]
\centering
	 \subfigure[]{\label{fig:Hpoll1}
    \includegraphics[width=0.48\linewidth]{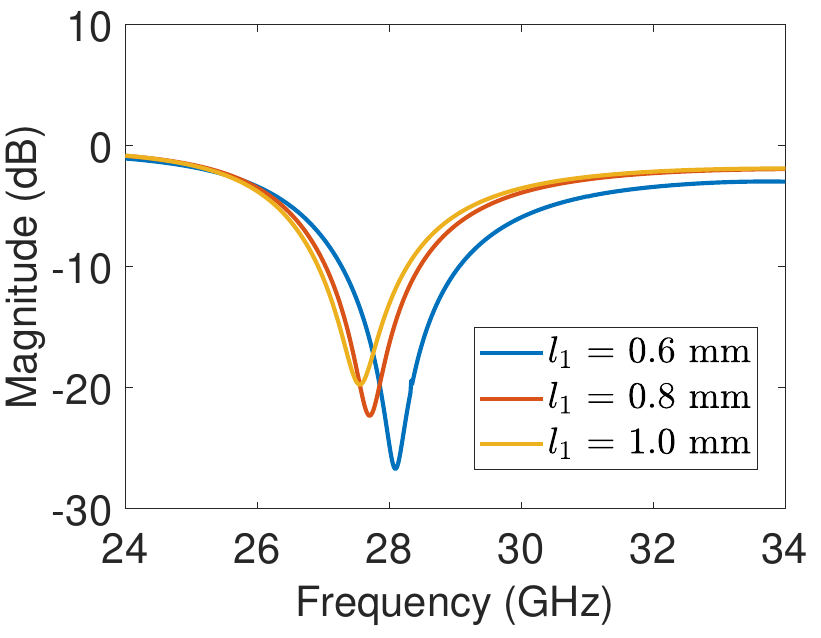}}
	 \subfigure[]{\label{fig:Hpolle}
    \includegraphics[width=0.48\linewidth]{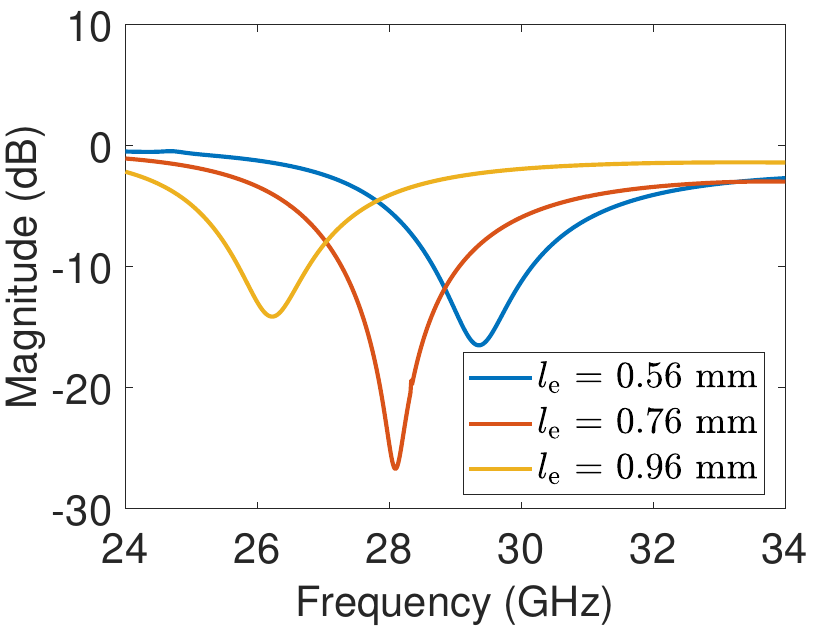}}
\caption{Reflection coefficients of the H-pol omnidirectional antenna (a) at $l_1 = 0.60$, $0.80$ and $1.00$~mm and (b) at $l_{\rm e} = 0.56$, $0.76$ and $0.96$~mm.}
 \label{fig:Hpolparameter}
\end{figure}

To investigate the influence of design parameters on the antenna’s reflection coefficients, we varied $l_1$ from $0.60$ to $1.00$~mm while keeping the other parameters in TABLE~\ref{tab:Hpolparameter} unchanged. As shown in Fig.~\ref{fig:Hpoll1}, this resulted in only a minor frequency shift. In contrast, changing $l_{\rm e}$ from $0.56$ to $0.96$~mm caused a significant resonance shift from $28$ to $26$~GHz. These observations indicate that the resonance frequency is primarily determined by the balun structure rather than the dipole elements.
 
\subsection{Antenna Fabrication}
\label{sec:hpolfabrication}
\begin{figure}[!ht]
    \centering
	\includegraphics[width=0.9\linewidth]{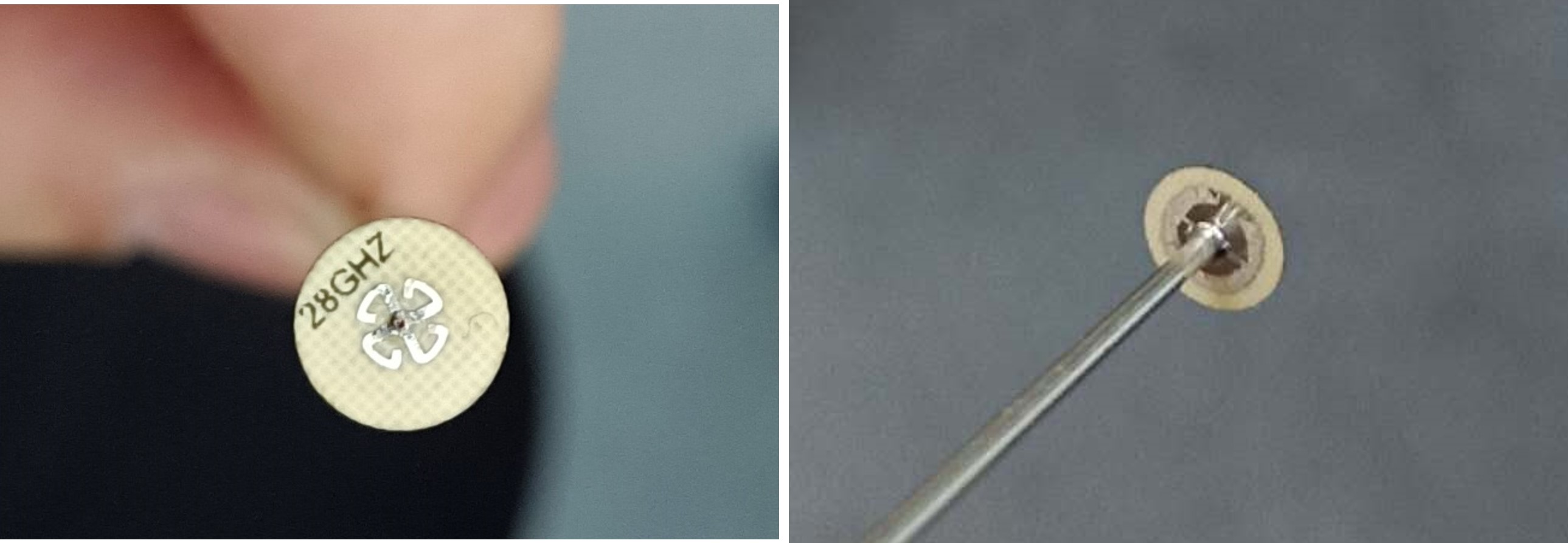}
	\caption{Top and bottom views of the H-pol antenna prototype.}
	\label{fig:Hpolprototype}
\end{figure}

The PCBs of the antenna were fabricated according to the parameters listed in Table~\ref{tab:Hpolparameter}, using a standard PCB process with a manufacturing tolerance of less than $1~\rm mil$. A semi-rigid coaxial cable features an open end with a bare inner conductor, while the other end is terminated with an SMA connector. Although SMA connectors are recommended only up to 18~GHz, they still offer acceptable performance to 40~GHz. 
The PCB outline was laser-cut to ensure high dimensional accuracy. The cable and antenna were soldered while a mechanical supporter maintained a semi-rigid cable and the PCB perpendicular to each other. The fabricated antenna prototype is shown in Fig.~\ref{fig:Hpolprototype}.

\subsection{Antenna Measurement}
\label{sec:hpolmeasurement}
\begin{figure}[!ht]
\centering
	 \subfigure[]{\label{fig:Hpolsparmmea}
    \includegraphics[width=0.49\linewidth]{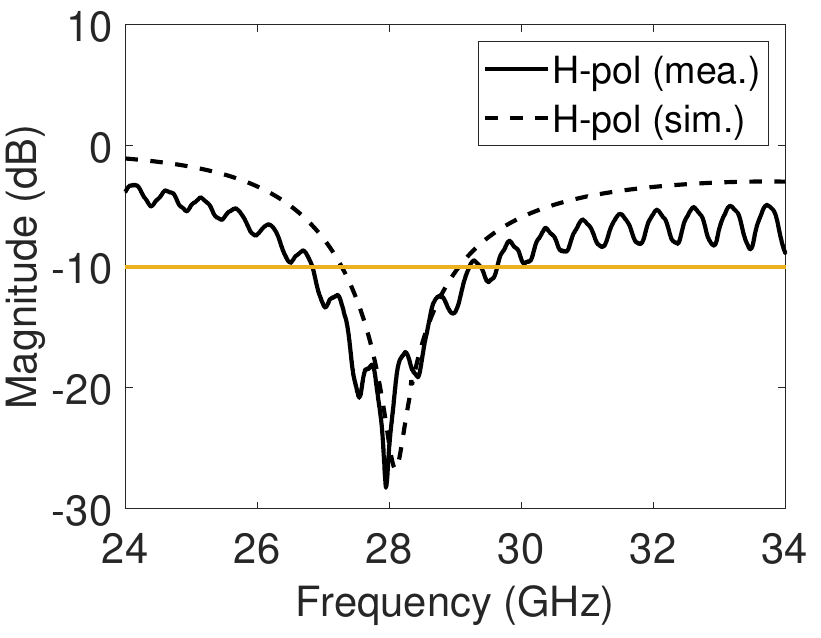}}
  \subfigure[]{\label{fig:HpolHplane}
    \includegraphics[width=0.435\linewidth]{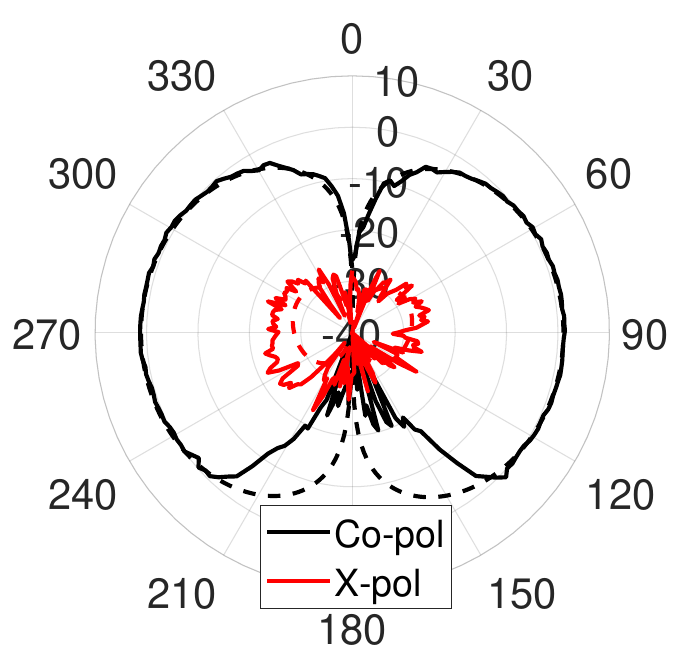}}
     \subfigure[]{\label{fig:HpolEplane}
    \includegraphics[width=0.435\linewidth]{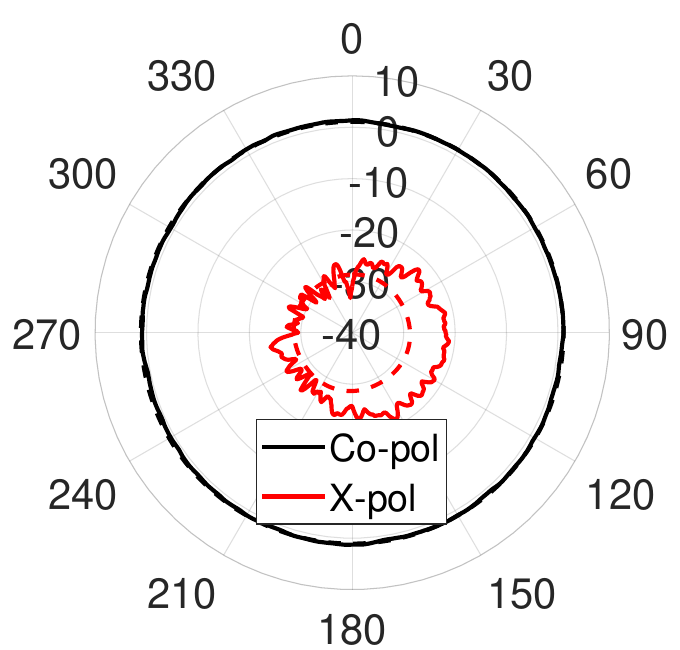}}
     \subfigure[]{\label{fig:HpolEplane2}
    \includegraphics[width=0.49\linewidth]{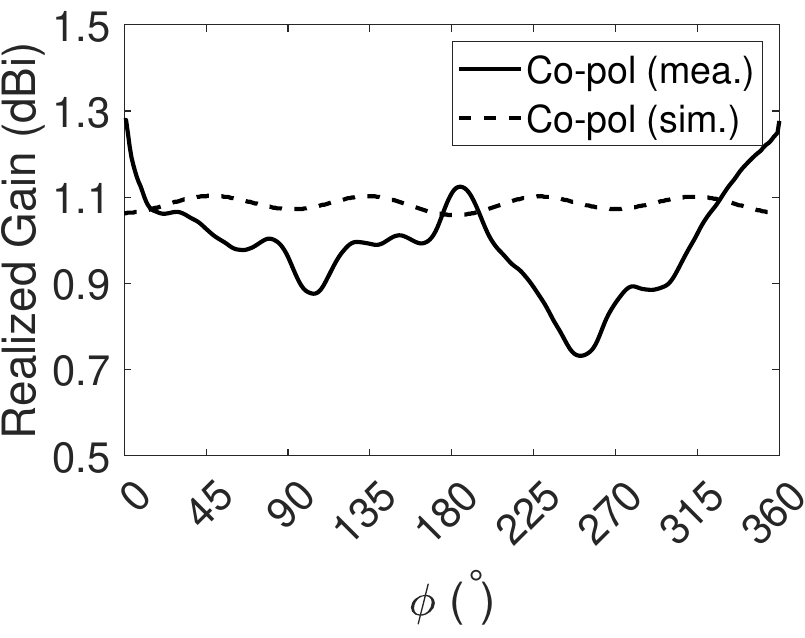}}
\caption{(a) Reflection coefficients, (b) H-plane radiation pattern, (c) E-plane radiation pattern, and (d) co-polarization on the E-plane of the H-pol omnidirectional antenna. Solid and dashed curves are from measurements and simulations; the measured radiation patterns are after de-embedding coaxial cable losses.}
	\label{fig:Hpolmea}
\end{figure}

The reflection coefficients of the H-pol antenna were measured by connecting its SMA connector to a 2.92~mm connector, where the reference plane of the vector network analyzer lies after calibrated. The measured reflection coefficients are shown in Fig.~\ref{fig:Hpolsparmmea}. The measured impedance bandwidth is slightly wider than the simulated result, reaching $2.2$~GHz. A resonance frequency shift of approximately $100$~MHz is observed. The ripples in the measurement are mainly attributed to a mismatch between the SMA and 2.92~mm connectors. The radiation pattern was evaluated using an anechoic spherical-scanning antenna measurement system. To de-embed the cable loss, transition cables with two SMA connectors of varying lengths were measured~\cite{Tuomela2025Design}. The cable loss was estimated from the transmission-coefficient difference between the two cables, yielding a value of $0.005$~dB/mm. The total cable loss, approximately $0.75$~dB for a 150~mm length, was compensated. The measured and simulated radiation patterns, shown in Fig.~\ref{fig:HpolHplane} and Fig.~\ref{fig:HpolEplane}, assume no cable loss in the simulation model. The simulated realized gain is $1.07$~dB, and the simulated radiation efficiency is $0.94$. The measured co-polarization pattern agrees well with the simulated patterns in both the E-plane and H-plane, with a maximum deviation of less than $0.5$~dB in the E-plane. A larger discrepancy in the H-plane is attributed to the absorber behind the antenna. The measured cross-polarization ratio is about $22$~dB, while the simulation shows a polarization ratio of about $30$~dB. This difference arises from environmental scattering and measurement uncertainties. Regarding gain variation, Fig.~\ref{fig:HpolEplane2} shows a measured variation of less than $0.6$~dB, whereas the simulated result is less than $0.07$~dB on the H-plane. This discrepancy is primarily due to measurement uncertainty and assembly errors in the antenna-cable interface.

\subsection{Comparisons with Existing Antennas}
Table~\ref{Hpoltable} illustrates a comparison between the proposed antenna and existing designs~\cite{Suzan2019Design,Zhou2021A,Liang2022Compact,Zhang2017Bandwidth,Peng2024Shared,Lin2019A,Ye2024Compact}. The proposed antenna offers comparable or wider impedance bandwidth to other designs. It has the smallest gain variation on the E-plane among those reported. The cross-polarization ratio is as high as those in existing antennas. Antenna assembly in our design is rather straightforward, reducing fabrication complexity relative to the intricate designs. These characteristics demonstrate the proposed antenna as a practical solution for a reference antenna at $28$~GHz.

\begin{table}[!ht]
	\begin{center}
		\caption{Comparisons with existing H-pol Omnidirectional antennas}
  \label{Hpoltable}
  \setlength{\tabcolsep}{1mm}{
		\begin{tabular}{cccccccc}
		\hline\hline Reference &Type& BW &Gain&$f_0$&GV&PR\\ \hline\hline
  ~\cite{Suzan2019Design}&Loop&$0.5$~GHz&$-2.5$~dBi&$28.0$~GHz&$2.1$~dB&N/A \\
   ~\cite{Zhou2021A}&MC&$0.09$~GHz&$7.1$~dBi&$2.45$~GHz&$2.3$~dB&$>20$~dB \\
  ~\cite{Liang2022Compact}&MC&$0.67$~GHz&$9.2$~dBi&$5.8$~GHz&$<3.0$~dB&$>38$~dB \\
  ~\cite{Zhang2017Bandwidth}&ECL&$2.30$~GHz&$0.5$~dBi&$2.73$~GHz&$2.2$~dB&$>23$~dB \\
  ~\cite{  Peng2024Shared}&ECL&$2.40$~GHz&$3.28$~dBi&$2.60$~GHz&$<1.5$~dB&$>16$~dB \\
    ~\cite{Lin2019A}&ECL&$4.0$~GHz&$1.22$~dBi&$38.0$~GHz&$2.0$~dB&N/A \\
    ~\cite{Ye2024Compact}&ECL&$6.0$~GHz&$7.2$~dBi&$27.0$~GHz&$<2.9$~dB&$>21$~dB \\\hline
    
  Ours&ECL&$2.2$~GHz&$1.35$~dBi&$28.0$~GHz&$<0.6$~dB&$>22$~dB\\\hline
   \end{tabular}}
	\end{center}
	BW is $-10$~dB impedance bandwidth; $f_0$ is the central working frequency; GV represents gain variation; PR is the polarization ratio of co-polarization to cross-polarization; N/A means `not available'; MC is equivalent magnetic current; ECL is equivalent electric current loop.
 \end{table}

\section{28 GHz V-pol Omnidirectional Antenna}
\label{sec:V-pol}
\subsection{Antenna Design}
\label{sec:V-poldesign}
The V-pol omnidirectional antenna primarily relies on an equivalent magnetic current loop or electric current source, which is `electric-magnetic' reciprocal to the H-pol omnidirectional antenna. A dipole antenna serves as a straightforward electric current source and hence a V-pol omnidirectional antenna. However, it requires an additional balun to connect the dipole arms to standard connectors, which can disrupt the symmetrical structure of a dipole~\cite{Suzan2019Design}.

Alternatively, equivalent magnetic current loops can be crafted using parallel circular patches with a low antenna profile and coaxial cable feeding~\cite{Wang2017Wideband}. This setup, however, poses challenges in impedance matching with a coaxial cable and requires supplementary structures~\cite{Lin2021Gain}. A biconical antenna, which amalgamates parallel circular patches and a dipole, can address these issues. It features progressively altered radiation arms with a central cable feeding, yet the challenge of mechanical stability at the transition remains.

At sub-6 GHz, low-permittivity sticks are often used to stabilize biconical antennas. This technique, however, is not suitable for the present case due to the difficulty in securing antennas with diminutive sticks and the increased sensitivity of antenna performance to manufacturing inaccuracies. Alternatively,~\cite{Zhang2019Wideband} and~\cite{Dobler2020An} employed lenses between conical structures as fixtures and to increase the gain;~\cite{Ratajczak2019Design} introduced a monopole-based feeding structure as a support, but both methods still entail high manufacturing complexity.

\begin{figure}[!ht]
\centering
	 \subfigure[]{\label{fig:designVpol1}
	\includegraphics[width=0.485\linewidth]{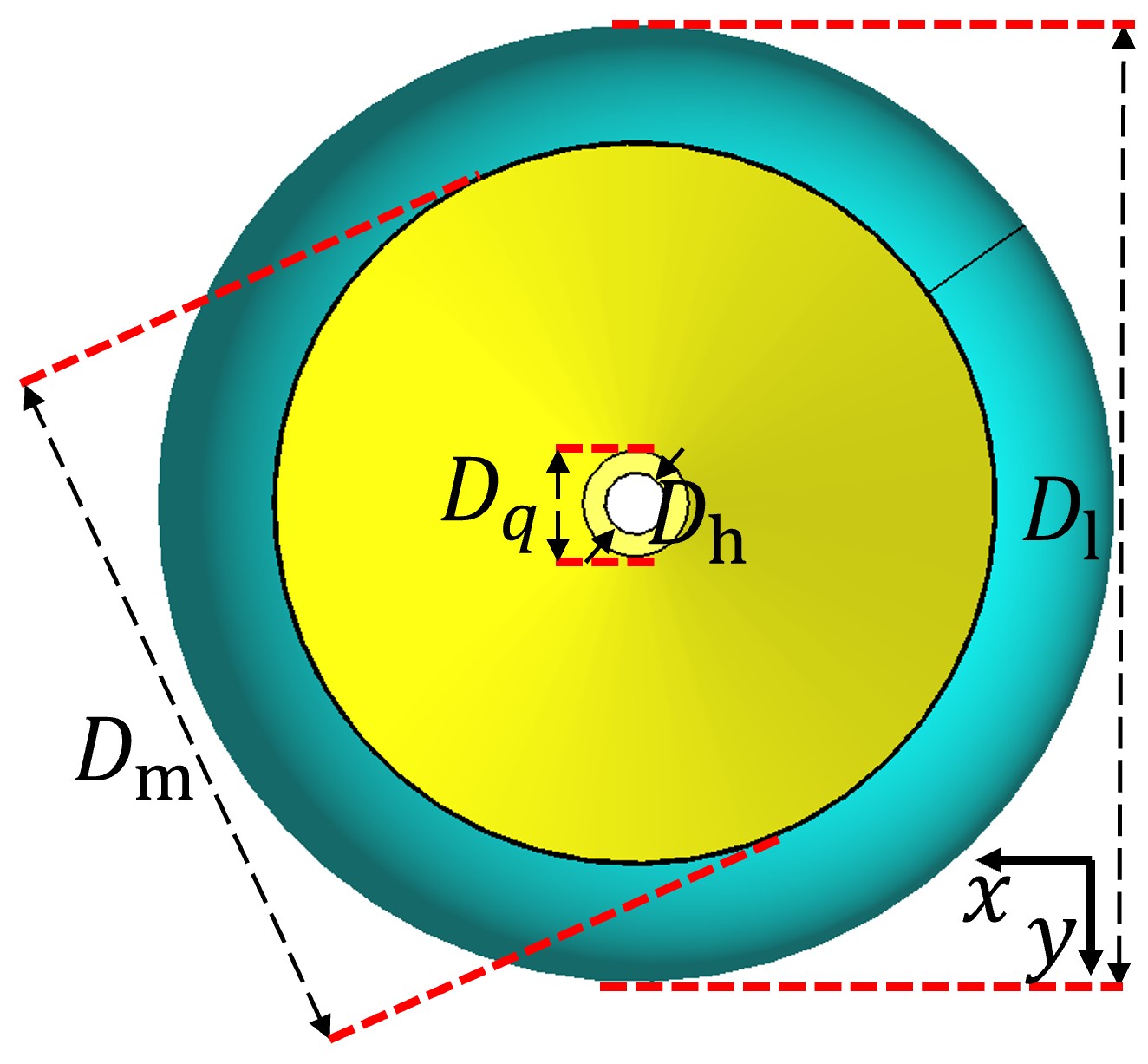}}
 	 \subfigure[]{\label{fig:designVpol2}
	\includegraphics[width=0.455\linewidth]{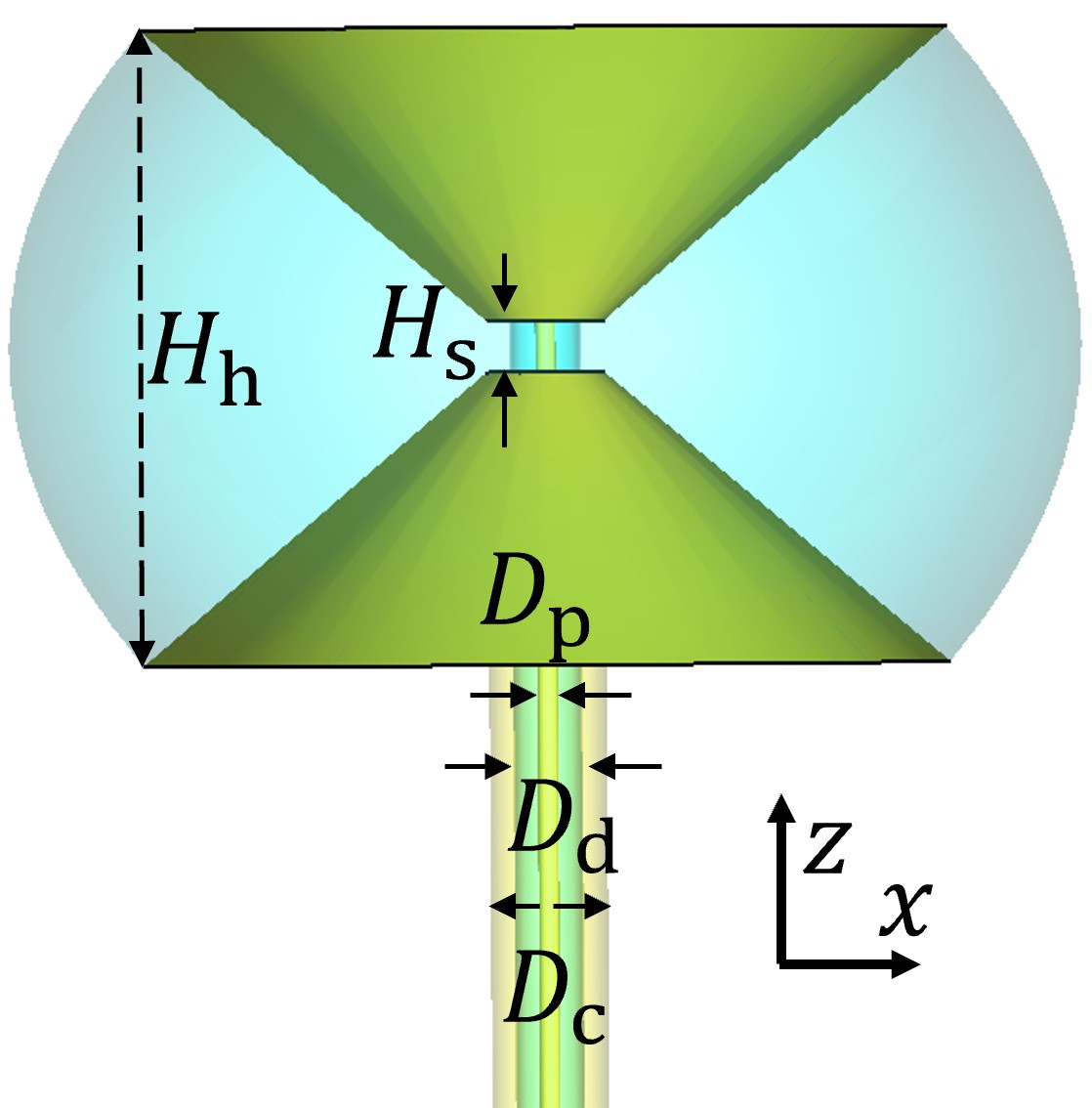}}
\caption{(a) Designed biconical structure of V-pol antenna; (b) designed V-pol antenna with feeding coaxial cable.}
	\label{fig:Vpoldesign}
\end{figure}

Our proposed antenna utilizes a biconical structure with a lens, and unlike the references~\cite{Zhang2019Wideband,Dobler2020An,Tuomela2025Design}, it operates without side lobes. Similar to the H-pol design, this antenna employs a thin coaxial cable as its feeding structure, as depicted in Fig.~\ref{fig:Vpoldesign}. The lens's outer curvature follows a parabolic equation to ensure radiation efficiency and support the metallic bicone as 
\begin{equation}
{z^2 = \frac{H_{\rm h}}{2(D_{\rm l}- D_{\rm m} )}x },
\label{eq:parabola}
\end{equation}
where the coordinate definition and $D_{\rm l}$, $D_{\rm m}$ and $H_{\rm h}$ are defined in Fig.~\ref{fig:Vpoldesign}. The lens material is an acrylonitrile butadiene styrene (ABS) plastic with a dielectric constant of $2.70$ and a loss tangent of $0.01$. Dimensions of the antenna are shown in Table~\ref{tab:Vpolparameter}.
\begin{table}[!ht]
	\begin{center}
		\caption{Dimensions of the proposed V-pol omnidirectional antenna}
		\label{tab:Vpolparameter}
        \begin{tabular}{llll}\hline\hline\multicolumn{4}{c}{V-pol omnidirectional antenna (unit: $\rm mm$)}\\\hline\hline
       $H_{\rm h} = 8.55$&$D_{\rm m} = 10.89$&$D_{\rm q} = 1.60$&$D_{\rm h} = 1.00$\\
        $D_{\rm d} = 0.94$&$D_{\rm l} = 14.46$&$D_{\rm p} = 0.94$&$H_{\rm s} = 1.00$\\
        $D_{\rm c} = 1.19$& & & \\\hline
		\end{tabular}
	\end{center}
	\end{table}

\begin{figure}[!ht]
\centering
	 \subfigure[]{\label{fig:Vpolsparm}
    \includegraphics[width=0.48\linewidth]{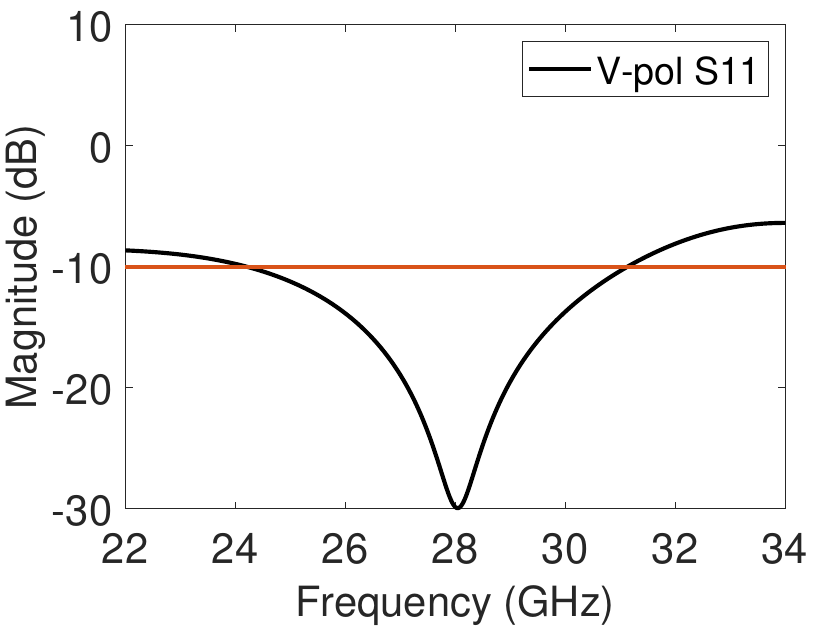}}
	 \subfigure[]{\label{fig:VpolEfield}
    \includegraphics[width=0.48\linewidth]{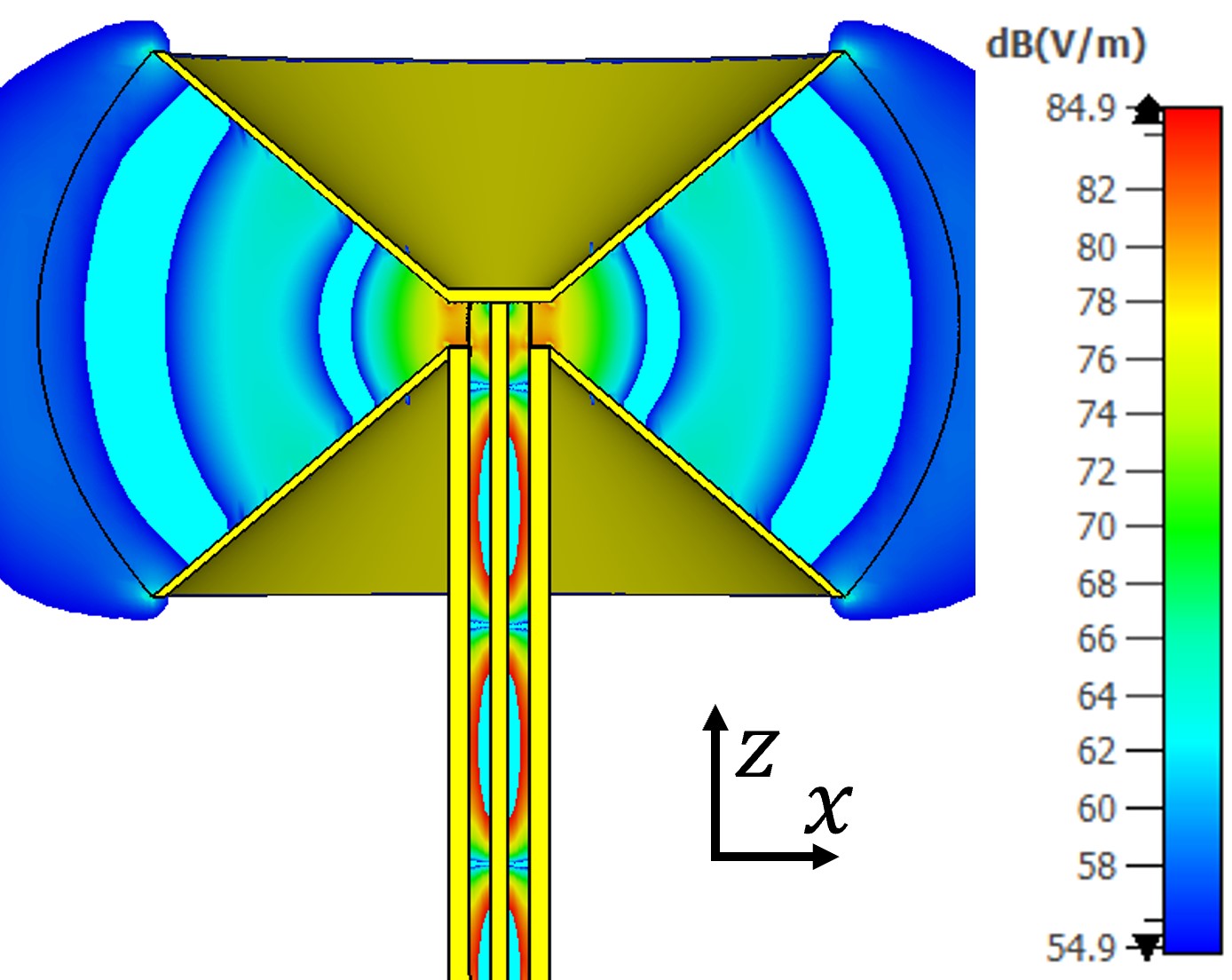}}
\caption{(a) Reflection coefficients, (b) E-Field on the substrate at $28$~GHz of the V-pol omnidirectional antenna.}
 \label{fig:Vpolsimulation}
\end{figure}

The $10$~mm coaxial cable is excited by a waveguide port, where the metal is $0.1$~mm-thick. Fig.~\ref{fig:Vpolsparm} shows that the antenna resonates at $28$~GHz with about $7.0$~GHz impedance bandwidth. Fig.~\ref{fig:VpolEfield} illustrates the E-fields inside the lens, indicating nearly no side lobe generated by the whole structure and a similar magnitude of the radiated wavefront.  

\begin{figure}[!ht]
\centering
	 \subfigure[]{\label{fig:VpolDmS}
    \includegraphics[width=0.49\linewidth]{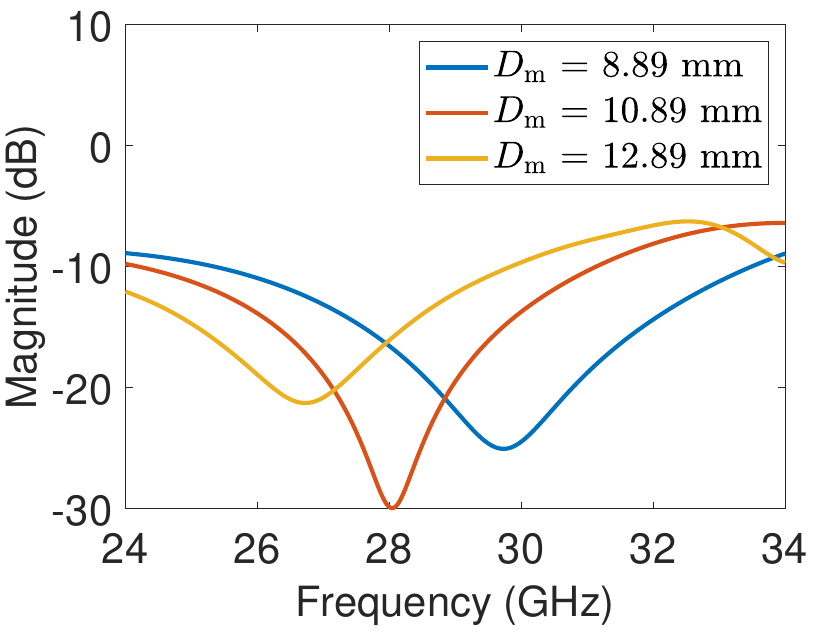}}
	 \subfigure[]{\label{fig:VpolDmP}
    \includegraphics[width=0.435\linewidth]{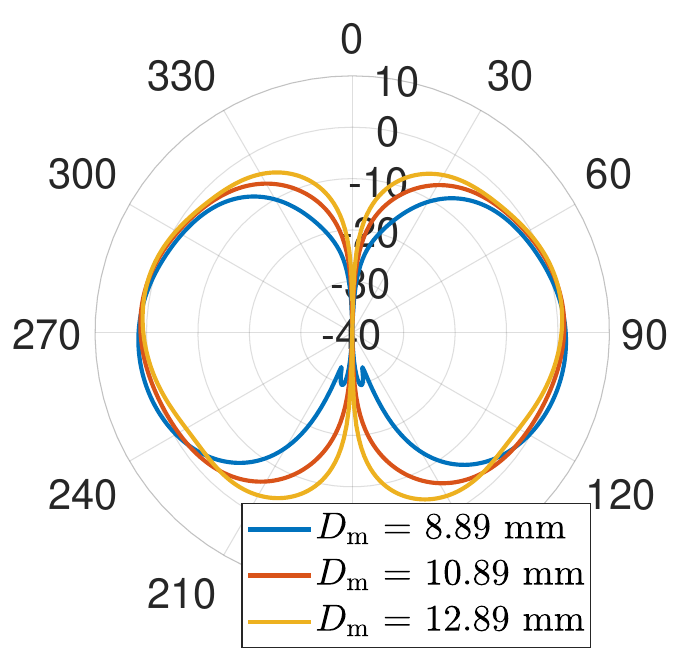}}
     \subfigure[]{\label{fig:VpolDlS}
    \includegraphics[width=0.49\linewidth]{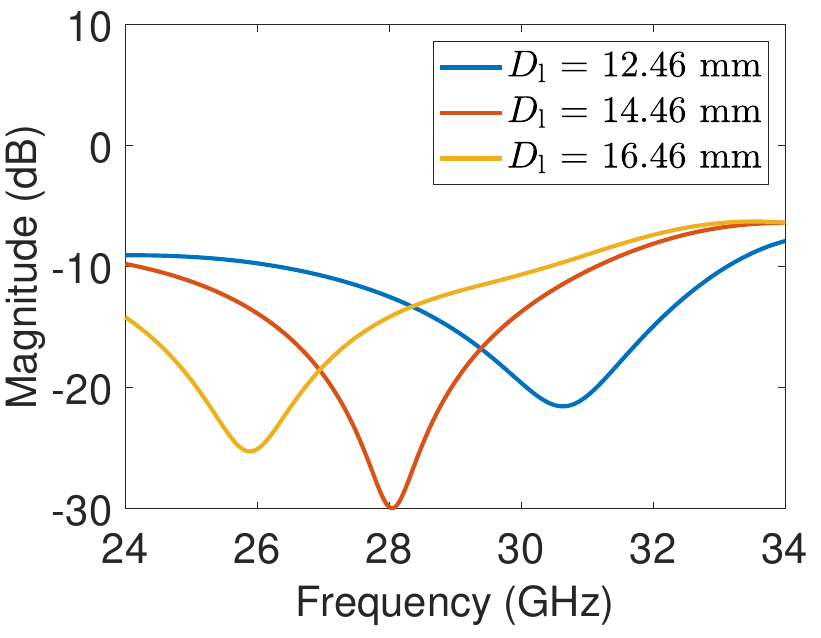}}
	 \subfigure[]{\label{fig:VpolDlP}
    \includegraphics[width=0.435\linewidth]{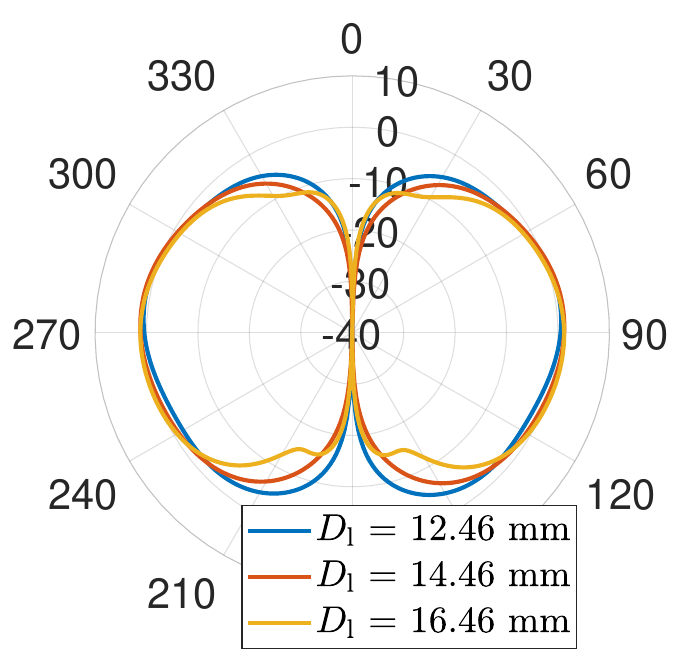}}
\caption{Reflection coefficients of the V-pol omnidirectional antenna (a) at $D_{\rm m} = 8.89$, $10.89$ and $12.89$~mm and (b) at $D_{\rm l} = 12.46$, $14.46$ and $16.46$~mm.}
 \label{fig:Vpolparameter}
\end{figure}

Parametric sweeps are performed for $D_{\rm m}$ ranging from $8.89$ to $12.89$~mm while keeping the other parameters in Table~\ref{tab:Vpolparameter} intact. There is a resonance frequency shift from $30$ to $26$~GHz as shown in Fig.~\ref{fig:VpolDmS}, while the beam becomes wider. On the contrary, when we change $D_{\rm l}$ from $12.46$ to $16.46$~mm, the beam becomes narrower while the resonance frequency changes from $31$ to $26$~GHz as shown in Fig.~\ref{fig:VpolDlS}. These two parameters can be used to make the antenna work at the target frequencies and beam width.

\subsection{Antenna Fabrication}
\label{sec:Vpolfabrication}
Computer numerical control (CNC) was employed to manufacture the dielectric part of the antenna, achieving a surface roughness (Ra) of 1.6 $\mu$m. The antenna is constructed from the ABS plastic base, shaped according to the parameters detailed in Table~\ref{tab:Vpolparameter}. To ensure high accuracy for the metal edges, copper metallic paint was applied by machine~\cite{Tuomela2025Design}. The same coaxial cables used in H-pol antennas serve as the feeding structure.

\begin{figure}[!ht]
    \centering
	\includegraphics[width=0.99\linewidth]{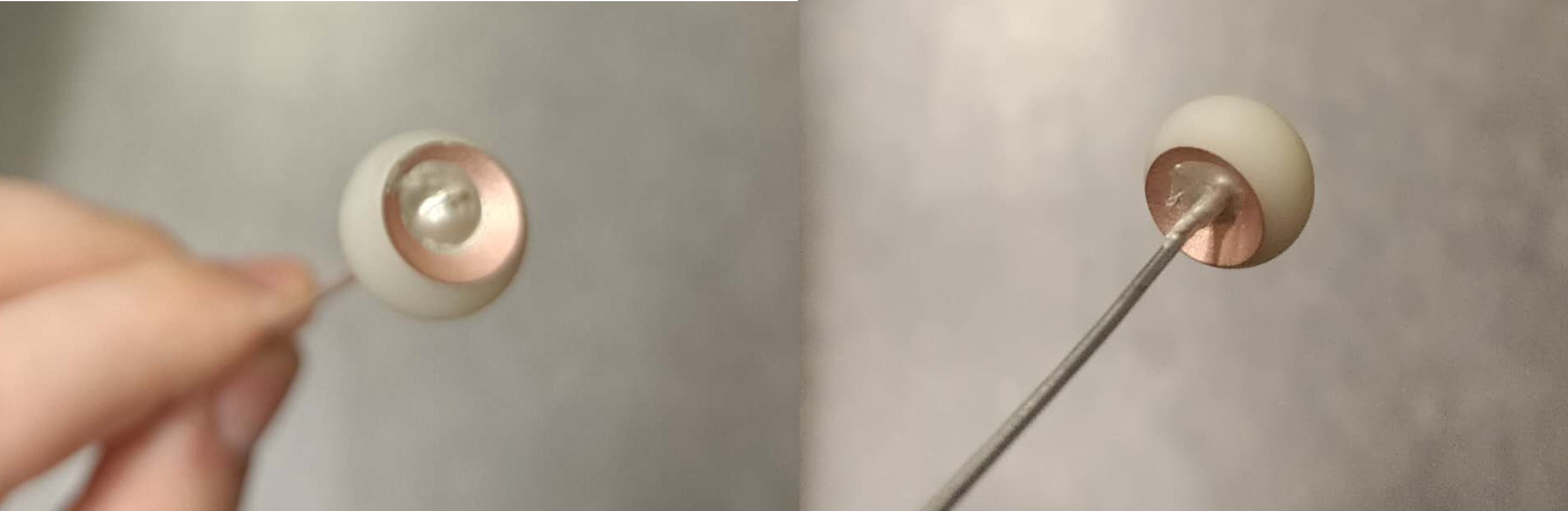}
	\caption{The top and bottom views of the V-pol antenna prototype.}
	\label{fig:Vpolprototype}
\end{figure}

Given that ABS is not tolerant to high temperatures, conventional tin solders were not used for connections between the cable and antenna. Instead, room-temperature solder paste was utilized~\cite{Tuomela2025Design}. To expedite its hardening, a heat gun was applied at the appropriate temperature, with completion achieved after approximately 30 minutes. During soldering, a simple fixture was employed to maintain the perpendicular alignment of the cable with the antenna~\cite{Tuomela2025Design}. The resulting antenna prototype is showcased in Fig.~\ref{fig:Vpolprototype}.

\subsection{Antenna Measurement}
\label{sec:Vpolmeasurement}
\begin{figure}[!ht]
\centering
	 \subfigure[]{\label{fig:Vpolsparmmea}
    \includegraphics[width=0.49\linewidth]{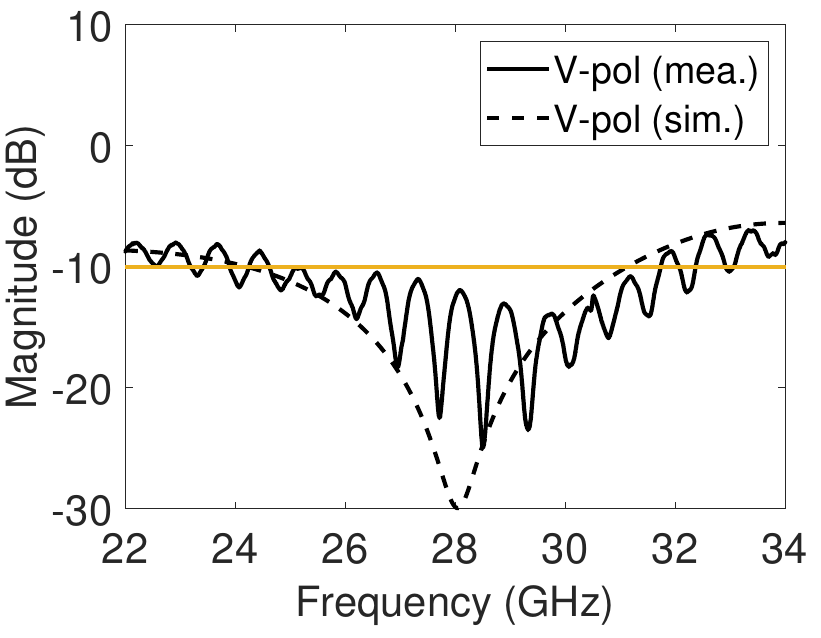}}
  \subfigure[]{\label{fig:VpolEplane}
    \includegraphics[width=0.435\linewidth]{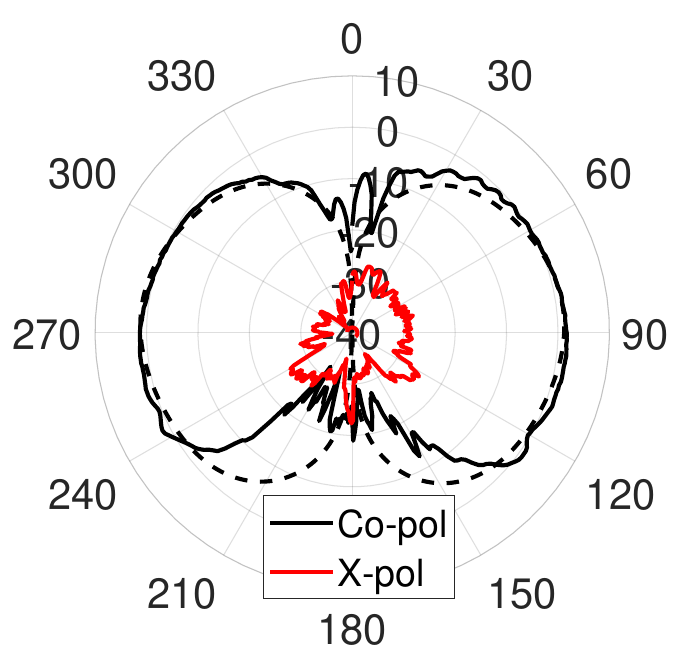}}
     \subfigure[]{\label{fig:VpolHplane}
    \includegraphics[width=0.435\linewidth]{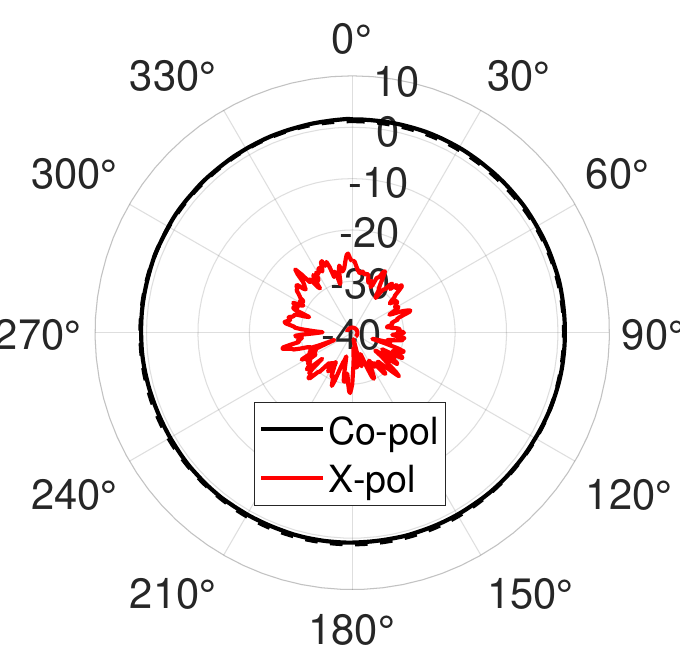}}
     \subfigure[]{\label{fig:VpolHplane2}
    \includegraphics[width=0.49\linewidth]{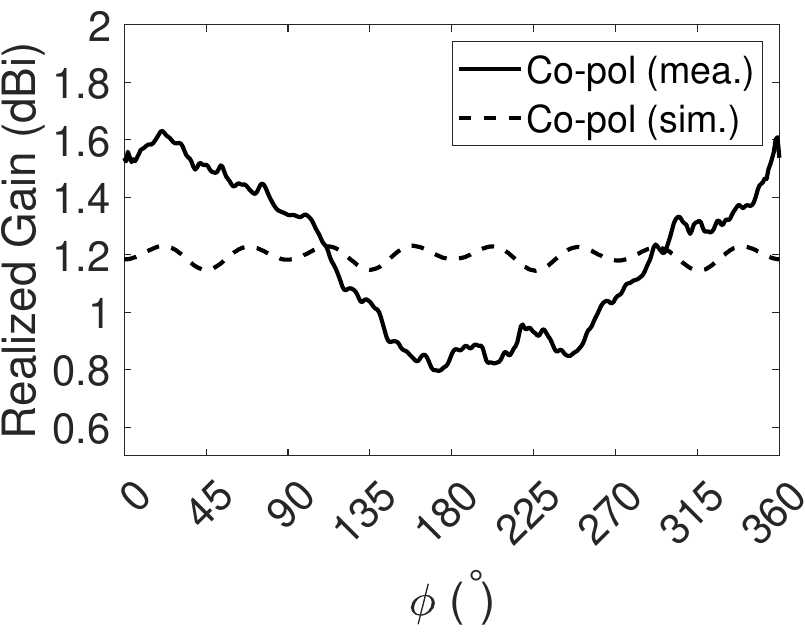}}
\caption{(a) Reflection coefficients, (b) E-plane radiation pattern, (c) H-plane radiation pattern, and (d) co-polarization on the H-plane of the V-pol omnidirectional antenna. Solid curves represent measured data; dash curves represent simulated data; the measured radiation patterns are already compensated by coaxial cables.}
	\label{fig:Vpolmea}
\end{figure}

The V-pol antenna was measured in the same manner as the H-pol antenna. Reflection coefficients shown in Fig.~\ref{fig:Vpolsparmmea} indicate that the measured impedance bandwidth is similar to the simulated curve, reaching $7.5$~GHz. The resonance frequency shifts by about $500$~MHz. The ripples are mainly because of impedance mismatch at the SMA to 2.92 mm connector interface and at the contact between the outer shell of the cable and the antenna's conical arm. Radiation patterns are shown in Fig.~\ref{fig:VpolEplane} and~\ref{fig:VpolHplane}, where the estimated loss of $150$~mm-cable in the measurement was de-embedded. The simulated realized gain of the antenna is about $1.20$~dB, and the simulated radiation efficiency is $0.76$ due to the ABS's dielectric loss. The measured co-polarization component is similar to the simulated curves, both for the E- and H-plane. The maximum difference is about $0.5$~dB on the H-plane, while the difference on the E-plane is due to the absorber at the back of the antenna. The cross-polarization ratio is around $25$~dB in the measurement, while the simulated one is around $40$~dB. This difference is because of scattering in the anechoic chamber and measurement uncertainties. Finally, the measured and simulated gains of the antenna on the H-plane are shown in Fig.~\ref{fig:VpolHplane2}. Their variation over azimuth angles are $0.8$~dB and $0.09$~dB. The larger variation in the measurement is due to measurement uncertainties and antenna-cable assembly errors.

\subsection{Comparisons with Existing Antennas}
Table~\ref{Hpoltable} shows the performance comparisons among the proposed antenna and other existing antennas. 
The proposed antenna has the lowest measured gain variation on the H-plane. The cross-polarization discrimination is close to the other antennas. In addition, the proposed antenna is easy to assemble with cables, showing a low fabrication complexity compared with the same antenna type proposed in~\cite{Zhang2019Wideband,Dobler2020An,Ratajczak2019Design}.

\begin{table}[!ht]
	\begin{center}
		\caption{Comparisons with existing V-pol Omnidirectional antennas}
  \label{Vpoltable}
  \setlength{\tabcolsep}{1mm}{
		\begin{tabular}{cccccccc}
		\hline\hline Reference &Type& BW &Gain&$f_0$&GV&PR\\ \hline\hline
  ~\cite{Suzan2019Design}&Dipole&$6.5$~GHz&$1.5$~dBi&$28.0$~GHz&$2$~dB&N/A \\
~\cite{Wang2017Wideband}$^*$&MCL&$8.2$~GHz&$4.5$~dBi&$30.5$~GHz&$0.5$~dB&$30$~dB \\
        ~\cite{Lin2021Gain}&MCL&$0.7$~GHz&$4.11$~dBi&$5.0$~GHz&N/A&N/A \\
                   ~\cite{Zhai2025}&MCL&$1.0$~GHz&$6.5$~dBi&$3.4$~GHz&$1.5$~dB&$20$~dB\\
   ~\cite{Zhang2019Wideband}$^*$&Bi-Lens&$6.0$~GHz&$11.0$~dBi&$27$~GHz&N/A&N/A \\
  ~\cite{Dobler2020An}&Bi-Lens&$>35$~GHz&$\sim 4$~dBi&$90$~GHz&$3.0$~dB&$10$~dB \\
          ~\cite{Ratajczak2019Design}$^*$&Bi-Lens&$7$~GHz&$4.5$~dBi&$60$~GHz&$4$~dB&N/A \\
        ~\cite{Tuomela2025Design}&Bi-Lens&$12$~GHz&$1.4$~dBi&$28.0$~GHz&$1.3$~dB&N/A \\
          ~\cite{Tang2024A}&Bicone&$25$~GHz&$5.0$~dBi&$62.5$~GHz&N/A&$>25$~dB \\\hline
  Ours&Bi-Lens&$7.5$~GHz&$1.2$~dBi&$28.0$~GHz&$0.8$~dB&$25$~dB\\\hline
   \end{tabular}}
	\end{center}
	BW is $-10$~dB impedance bandwidth; $f_0$ is the central working frequency; GV represents the gain variation; PR is the polarization ratio of co-polarization to cross-polarization; N/A means `not available'; MCL is an equivalent magnetic current loop; Bi-Lens is the biconical antenna with a lens; $^*$ means only simulation results.
 \end{table}

\section{Reference Antenna Module}
\label{sec:HVpolarray}
\subsection{V-pol and H-pol Antenna Separations}
\begin{figure}[!ht]
\centering
	\subfigure[]{\label{fig:Vpolpair}
    \includegraphics[width=0.355\linewidth]{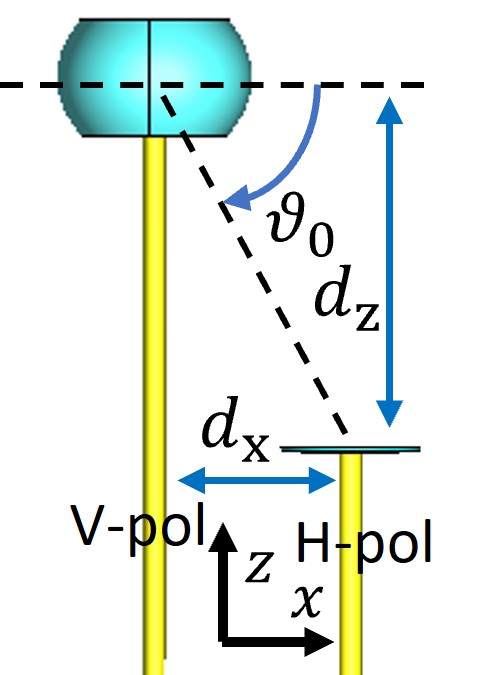}}
 \subfigure[]{\label{fig:Hpolpair}
    \includegraphics[width=0.325\linewidth]{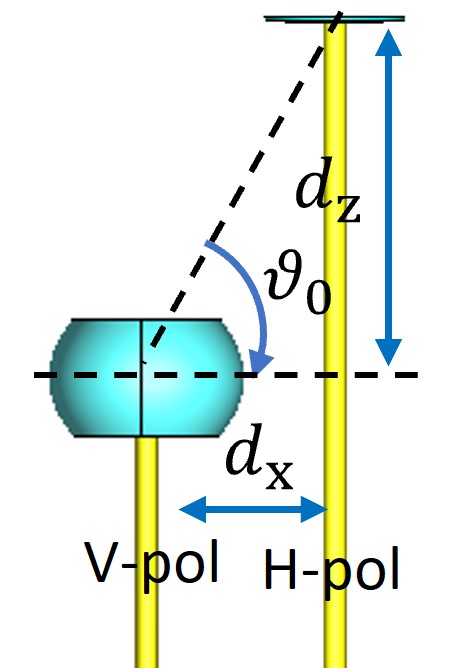}}
\caption{(a) Measurement setup of the reference antenna module with (a) dominant V-pol component ($d_{\rm z}>0$) and (b) dominant H-pol component ($d_{\rm z}<0$).}
	\label{fig:pair}
\end{figure}

For wireless channel measurements, it is ideal if the V- and H-pol elements of the omnidirectional antenna module are co-located. In practice, the two elements should be situated close to each other so that they see similar radio environments and fulfill the wide-sense stationarity~\cite[Section 6.4.1]{molisch2022wireless}. However, positioning these two antennas closely results in mutual interference that affects their radiation patterns. As illustrated in Fig.~\ref{fig:pair}, antennas can be positioned with varying vertical separations $d_{\rm z}$ and horizontal distances $d_{\rm x}$. 

By running full-wave simulations, we derive the gain variations across different values of $d_{\rm x}$ and $d_{\rm z}$ as displayed in Table~\ref{VHtable}. When both antennas are at the same height, i.e., $d_{\rm z} = 0$~mm, the gain varies significantly on the horizontal plane. To mitigate the interference, employing a vertical separation is beneficial to reduce scattering among their radiation patterns within low elevation angles. Thanks to the thin nature of applied coaxial cables, scattering from them is minimal. 
Gain variations of the V-pol antenna remain minimal at $d_{\rm z}>0$ due to no cable at its horizontal plane, whereas those of the H-pol antenna are reduced at $d_{\rm z}<0$. Here, $d_{\rm z}>0$ means V-pol antenna is above the H-pol antenna and vice versa. Therefore, for radio environments where the V-pol field is stronger, the setup in Fig.~\ref{fig:Vpolpair} is suitable. Alternatively, the arrangement in Fig.~\ref{fig:Hpolpair} is appropriate for environments dominated by the H-pol fields.

\begin{table}[!ht]
	\begin{center}
		\caption{Gain variations with different separations}
  \label{VHtable}
  \setlength{\tabcolsep}{1mm}{
		\begin{tabular}{cccccccc}
		\hline\hline unit:  &$d_{\rm x}=5$ &$d_{\rm x}=5$ &$d_{\rm x}=5$ &$d_{\rm x}=5$ &$d_{\rm x}=7$ &$d_{\rm x}=9$\\
                        (mm) &$d_{\rm z}=10$&$d_{\rm z}=15$&$d_{\rm z}=20$&$d_{\rm z}=25$&$d_{\rm z}=25$&$d_{\rm z}=25$\\
        \hline
                    V-pol&$0.6$~dB&$0.4$~dB&$0.4$~dB&$0.3$~dB&$0.3$~dB&$0.4$~dB \\
                H-pol&$2.8$~dB&$1.8$~dB&$1.5$~dB&$1.2$~dB&$1.4$~dB&$1.1$~dB \\
		\hline\hline
        unit:  &$d_{\rm x}=12$ &$d_{\rm x}=14$ &$d_{\rm x}=15$ &$d_{\rm x}=15$ &$d_{\rm x}=10$ &$d_{\rm x}=15$\\
                        (mm) &$d_{\rm z}=25$&$d_{\rm z}=25$&$d_{\rm z}=25$&$d_{\rm z}=-25$&$d_{\rm z}=-25$&$d_{\rm z}=0$\\
        \hline
                    V-pol&$0.3$~dB&$0.4$~dB&$0.6$~dB&$4.3$~dB&$4.2$~dB&$2.2$~dB \\
                    H-pol&$1.3$~dB&$1.4$~dB&$1.6$~dB&$1.5$~dB&$0.6$~dB&$13.4$~dB \\
\hline\hline
   \end{tabular}}
	\end{center}
 \end{table}

Compared to the configuration with $d_{\rm z}<0$, smaller gain variations are observed when $d_{\rm z}>0$. In instances where channel characteristics are not fully understood, the setup illustrated in Fig.~\ref{fig:Vpolpair} is recommended. The configuration with $d_{\rm x}=9$~mm and $d_{\rm z}=25$~mm is shown to offer sufficiently low gain variations according to Table~\ref{VHtable}. By separating the antennas based on these values, we can achieve reduced interference and hence low gain variations.

\subsection{Antenna Module Pattern Characterization}
\begin{figure}[!ht]
\centering
	 \subfigure[]{\label{fig:Vhpolpair}
    \includegraphics[width=0.7\linewidth]{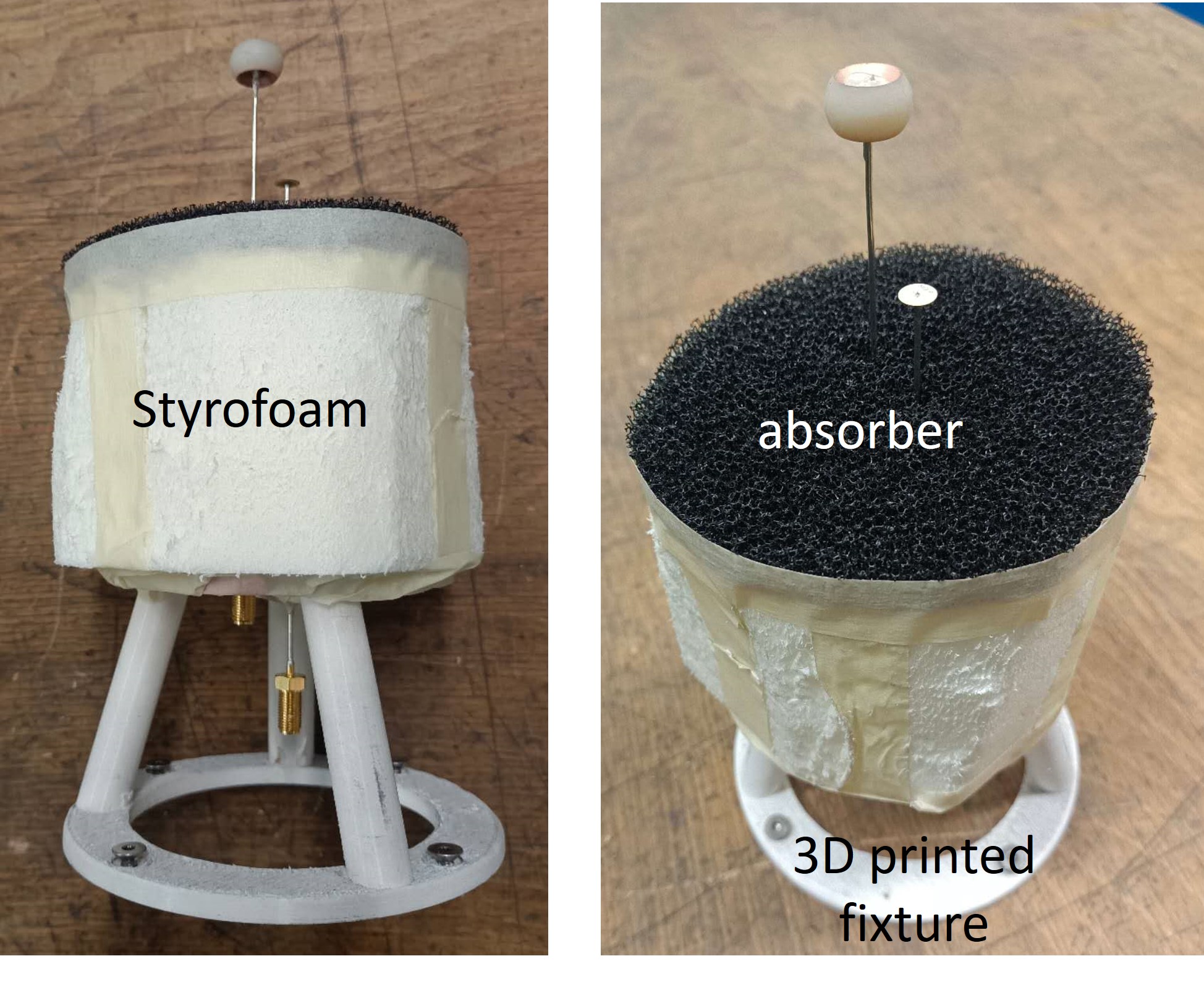}}
  \subfigure[]{\label{fig:pairmeasurement}
   \includegraphics[width=0.7\linewidth]{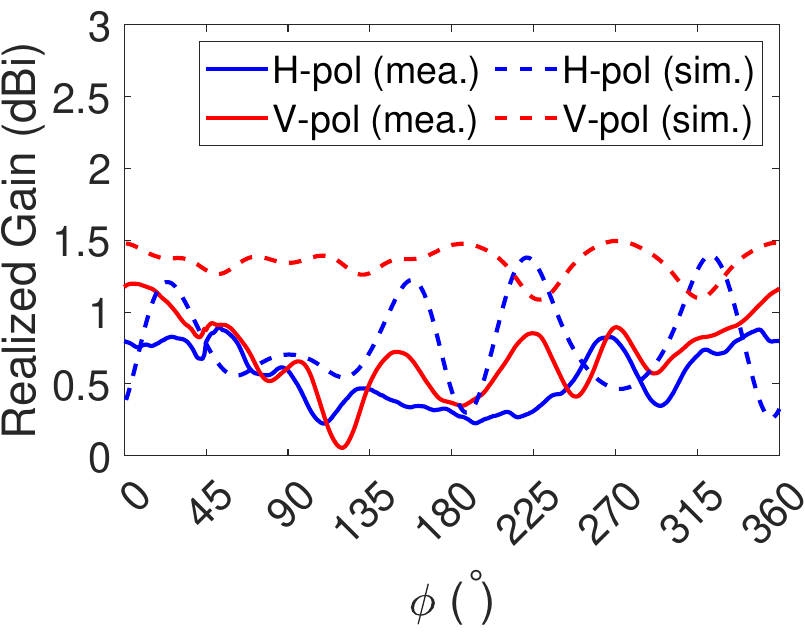}}
\caption{(a) Different views of V-pol and H-pol reference antenna module; (b) simulated and measured radiation patterns of the setup with $d_x \approx 10$~mm, and $d_{\rm z} \approx 22$~mm for V-pol and H-pol reference antennas at $\theta=90^\circ$}
	\label{fig:pairmea}
\end{figure}

The reference antenna module, consisting of a pair of H-pol and V-pol elements, was fabricated as depicted in Fig.~\ref{fig:Vhpolpair}, utilizing 3D printed support structures with the antennas fixed by absorber-covered styrofoam. Due to factors such as placement errors, soldering deficiencies, and the flexibility of semi-rigid cables, replicating a setup identical to full-wave simulations proved impractical. Initial settings of $d_x \approx 9$~mm and $d_{\rm z} \approx 25$~mm were adjusted through a trial-and-error to find the smallest gain variation, concluded with $d_x \approx 10$~mm and $d_{\rm z} \approx 22$~mm.

The measured radiation patterns of the two antennas, accounting for a cable loss, are shown in Fig.~\ref{fig:pairmeasurement}. The simulated radiation patterns without cable loss are also shown. The measured gain variation for the V-pol antenna is $1.14$~dB due to scattering from the H-pol antenna, while the H-pol antenna achieves a gain variation of $0.70$~dB, similar to its standalone behavior as illustrated in Fig.~\ref{fig:HpolEplane2}. It can be seen that the measured gain variation is not worse than that observed in the simulations.

\subsection{Comparisons with Existing Dual-Polarized Omnidirectional Antennas}
Table~\ref{HVpoltable} shows the performance comparisons among the proposed antenna module and other dual-polarized omnidirectional antennas. The proposed antenna has the lowest measured gain variation on the H-plane. The cross-polarization discrimination is close to the best value of other antennas. In addition, the proposed antenna is easy to manufacture at mmWaves, showing a low fabrication complexity compared with others. 
\begin{table}[!ht]
	\begin{center}
		\caption{Comparisons with existing Dual-polarized Omnidirectional antennas}
  \label{HVpoltable}
  \setlength{\tabcolsep}{1mm}{
		\begin{tabular}{ccccccc}
		\hline\hline Ref &Types& OL-BW &$f_0$&H- / V-pol GV&ISo\\ \hline\hline
  ~\cite{Suzan2019Design}&Dipole / loop&$0.5$~GHz&$28$~GHz&$2.1$ / $2$~dB&N/A\\
~\cite{Zhai2024A}&Conical / MCL&$1.05$~GHz&$2.2$~GHz&$1.7$ / $0.7$~dB&$20$~dB \\
~\cite{Yan2024A}&Conical / MCL&$8.2$~GHz&$6.5$~GHz&$3.8$ / $3$~dB&$20$~dB \\
~\cite{Zhou2025A}$^*$&Conical / MCL&$2.8$~GHz&$2.3$~GHz&$3$ / $1.6$~dB&$35$~dB\\
   ~\cite{Ke2025Dual}&Monopole / MCL&$2.2$~GHz&$2.7$~GHz&$2.4$~dB&$25$~dB\\\hline
  Ours&Bi-Lens-MCL&$2.2$~GHz&$28$~GHz&$0.7$ / $1.14$~dB&$32$~dB\\\hline
   \end{tabular}}
	\end{center}
	Ref: reference; Overlapping BW: H-pol and V-pol overlapping $-10$~dB impedance bandwidth; $f_0$ is the central working frequency; GV represents the gain variation; N/A means `not available'; MCL is an equivalent magnetic current loop; Bi-Lens is the biconical antenna with a lens; ISo: V-pol and H-pol Isolation; $^*$ means only simulation results.
 \end{table}
 
\section{Conclusions}
\label{sec:conclusion}
In this manuscript, we have introduced a dual-polarized, compact mmWave omnidirectional antenna module, encompassing the design, fabrication, and measurement processes. The proposed V-pol and H-pol antenna elements have gain variations of 0.8~dB and 0.6~dB, almost identical realized gain on the horizontal plane, and effectively address the challenges associated with manufacturing complexity. When integrated in a compact volume, it serves as a reference antenna module that shows gain variations of 1.14~dB and 0.7~dB for the two polarizations and hence is applicable to handset antenna testing in multipath environments.

\bibliographystyle{IEEEtran}
\footnotesize
\bibliography{refMPAC}

\end{document}